\documentclass[preprint,12pt]{elsarticle}

\usepackage{amssymb}
\usepackage{amsmath}

\usepackage{xurl}

\begin{document}

\begin{frontmatter}

\title{A longitudinal analysis of misinformation, polarization and toxicity on Bluesky after its public launch}

\author[1]{Gianluca Nogara\corref{cor1}}
\ead{gianluca.nogara@supsi.ch}
\author[1]{Erfan Samieyan Sahneh}
\author[2]{Matthew R. DeVerna}
\author[2]{Nick Liu}
\author[3]{Luca Luceri}
\author[2]{Filippo Menczer}
\author[4]{Francesco Pierri}
\author[1]{Silvia Giordano}

%
\affiliation[1]{organization={ISIN - DTI, SUPSI},
city={Lugano},
country={Switzerland}}

\affiliation[2]{organization={Observatory on Social Media, Indiana University},
city={Bloomington},
country={USA}}

 \affiliation[3]{organization={USC Information Sciences Institute},
city={Los Angeles, California},
country={USA}}

\affiliation[4]{organization={Dip. Elettronica, Informazione e Bioingegneria, Politecnico di Milano},
city={Milano},
country={Italy}}

\begin{abstract}
Bluesky is a decentralized, Twitter-like social media platform that has rapidly gained popularity. Following an invite-only phase, it officially opened to the public on February 6th, 2024, leading to a significant expansion of its user base. In this paper, we present a longitudinal analysis of user activity in the two months surrounding its public launch, examining how the platform evolved due to this rapid growth.
Our analysis reveals that Bluesky exhibits an activity distribution comparable to more established social platforms, yet it features a higher volume of original content relative to reshared posts and maintains low toxicity levels. We further investigate the political leanings of its user base, misinformation dynamics, and engagement in harmful conversations. Our findings indicate that Bluesky users predominantly lean left politically and tend to share high-credibility sources. After the platform's public launch, an influx of new users—particularly those posting in English and Japanese—contributed to a surge in activity. Among them, several accounts displayed suspicious behaviors, such as mass-following users and sharing content from low-credibility news sources. Some of these accounts have already been flagged as spam or suspended, suggesting that Bluesky's moderation efforts have been effective. 
\end{abstract}


\end{frontmatter}

\section{Introduction}

Decentralized social media platforms have emerged as alternatives to traditional, centralized networks, offering users greater control over their data, content moderation, and online interactions. Unlike mainstream platforms, which rely on centralized governance and proprietary algorithms, decentralized platforms distribute authority across independent servers or protocols, reducing single points of failure and promoting transparency~\cite{bono2024exploration,Raman2019ChallengesIT,Zignani2018FollowT,Kleppmann2024ATProtocol}.

Bluesky Social\footnote{\url{https://bsky.social/about}} is a novel decentralized social media platform for microblogging based in the United States. Founded by Jack Dorsey as a Twitter-like alternative, Bluesky aims to provide a more open and user-controlled social networking experience. Originally based on an invite-only subscription, the platform officially opened to the public on February 6th, 2024~\cite{Silberling:Bluesky-opening}. With only a few thousand users active in January 2024, the platform registered more than one million new users on the first day of its opening~\cite{Bluesky2024}. Recent events, such as the banning of X/Twitter in Brazil and the 2024 U.S. presidential election, have contributed to a surge in new users on Bluesky, particularly among those who have grown distrustful of X and are seeking an alternative platform for political discussions and online discourse~\cite{lacava_2023_drivers, placido_2024_the, boyd_2024_from}. As a result, Bluesky has attracted millions of new users, solidifying its position as a viable alternative to mainstream platforms~\cite{butts_2024_social}.

As decentralized platforms like Bluesky gain traction, they present new opportunities and challenges in online social dynamics, user behavior, and moderation strategies. On the one hand, they empower users with greater control over their online interactions, reduce reliance on centralized authorities, and promote transparency in content moderation~\cite{Kleppmann2024ATProtocol}. On the other hand, the decentralized nature of these platforms raises concerns about the effectiveness of moderation, the spread of misinformation, and the potential for fragmented online communities with varying moderation standards~\cite{Berg2021Decentralized-moderation}.

Here, building upon the preliminary findings of our previous work~\cite{erfansamieyansahneh_2025_theDawn}, we conduct a comprehensive analysis of user activity on Bluesky following its public launch on February 6th, 2024. Specifically, we examine key characteristics of the platform, including temporal patterns of user engagement, the prevalence of different languages, and the structural properties of the follower network. We then investigate the estimated political leanings of Bluesky users and their relationship with sharing information from sources of varying reliability and participation in harmful conversations. Lastly, we analyze the largest communities on the platform and assess the extent and effectiveness of content moderation measures implemented during Bluesky’s initial months of global availability.
This work expands on our preliminary report~\cite{erfansamieyansahneh_2025_theDawn} by adding an analysis of source credibility, examining the trend of high and low credibility sources along the timeline, and extending the study of toxicity from English and Japanese to other languages in the data. We also build a resharing network by linking users who shared posts, which allows us to analyze communities based on the characteristics of the users. Finally, we leverage user status information to analyze how many users have been moderated by the platform or are no longer present. 


\section{Related work}

Distributed social networks aim for decentralization, allowing users to have more control and privacy. Early efforts like LifeSocial.KOM~\cite{Graffi2011LifeSocialKOMAS} and PeerSON~\cite{Buchegger2009PeerSoNPS} were based on the peer-to-peer model but faced challenges in performance and reliability.
This led to a shift toward server-based federated models like Mastodon~\cite{Raman2019ChallengesIT, Zignani2018FollowT}.
This approach balances flexibility and ease of use while maintaining some decentralization~\cite{bono2024exploration}.

Mastodon, created in 2016, is a free and open-source social media platform that allows users to create their own servers (``instances'') and connect with others across the globe.
It is a decentralized social network, meaning that it is not owned by a single entity, but rather a network of independent servers that are connected together. 
A number of studies have identified striking features that make up Mastodon's distinct ``fingerprint,'' distinguishing it from better-known online social networks~\cite{LaCava2021UnderstandingTG,bono2024exploration}.
Mastodon has, however, suffered from some natural pressures towards centralization, which can lead to potential points of failure~\cite{Raman2019ChallengesIT}.

To overcome this weakness, Bluesky developed its own decentralized AT protocol. 
Scalability, security, and ease of use make it an attractive option for building open and decentralized social media applications that prioritize user privacy and data security~\cite{Kleppmann2024ATProtocol}. 
Using standard web technologies and re-using existing data models from the Web 3.0 protocol family also contribute to its efficiency and reliability~\cite{edwards2023social}. 
Additionally, its federated networking model bolsters security by dispersing data across numerous servers, mitigating the risk of a single point of failure~\cite{Raman2019ChallengesIT}. 

Existing literature on Bluesky primarily focuses on general platform analysis, highlighting its features and overall structure. Several studies have delved into the political dynamics and polarization within social networks, focusing on user activity patterns and interactions~\cite{quelle2024bluesky}. These analyses reveal how user interactions often lead to the formation of ideological groupings, which can evolve into echo chambers. Additionally, the impact of these dynamics on political division within decentralized social networks has been thoroughly examined. Other research has explored the platform's federated model~\cite{failla2024m}, which allows users to curate their own experiences through customizable user feeds while maintaining interoperability between instances. 
Additional studies have focused on content analysis and user activities~\cite{balduf2024bluesky}. They have identified trends in user engagement and content virality, such as the types of posts that gain traction and the role of influencers in shaping discussions, highlighting differences from traditional centralized platforms. Finally, interactions were collected at the millisecond level by creating a multi-network temporal dataset~\cite{jeong_2024_descriptor} in order to perform an analysis of complex temporal dynamics such as community formation and social sanction patterns, i.e., user blocking.

Differently from previous work, our work focus on the evolution of the platform during its opening to the public to see if this has led to substantial changes in users or activities. Before the opening, the platform's controlled access reduced the risk of unauthorized or harmful usage. To study how this changed after the opening, we conducted an extensive longitudinal study of users and their activities before and after the opening.

\section{Methods}

\subsection{Data collection}
\label{data_collection}

We collected data using Bluesky's free and publicly available \textit{Firehose} endpoint, which grants developers real-time access to platform activities such as user posts, follows, and likes~\cite{bluesky_firehose,atProtocol_eventStream}. This endpoint ensures uninterrupted data collection by allowing reconnection and retrieval of up to 72 hours of past data in case of disruptions, ensuring comprehensive coverage during the observation period.

We employed the \texttt{dart} library from the AT protocol~\cite{BlueskyDart2024}, leveraging the Firehose endpoint\footnote{\url{https://atprotodart.com/docs/lexicons/com/atproto/sync/subscriberepos/}}~\cite{BlueskyFirehose2024}. 
This data collection process was performed through the use of an existing open-source project~\cite{BurghardtFirehose2024}. The process is both straightforward and flexible, as the AT protocol does not require user authentication for access.

\begin{figure}[!t]
\centerline{\includegraphics[width=\linewidth]{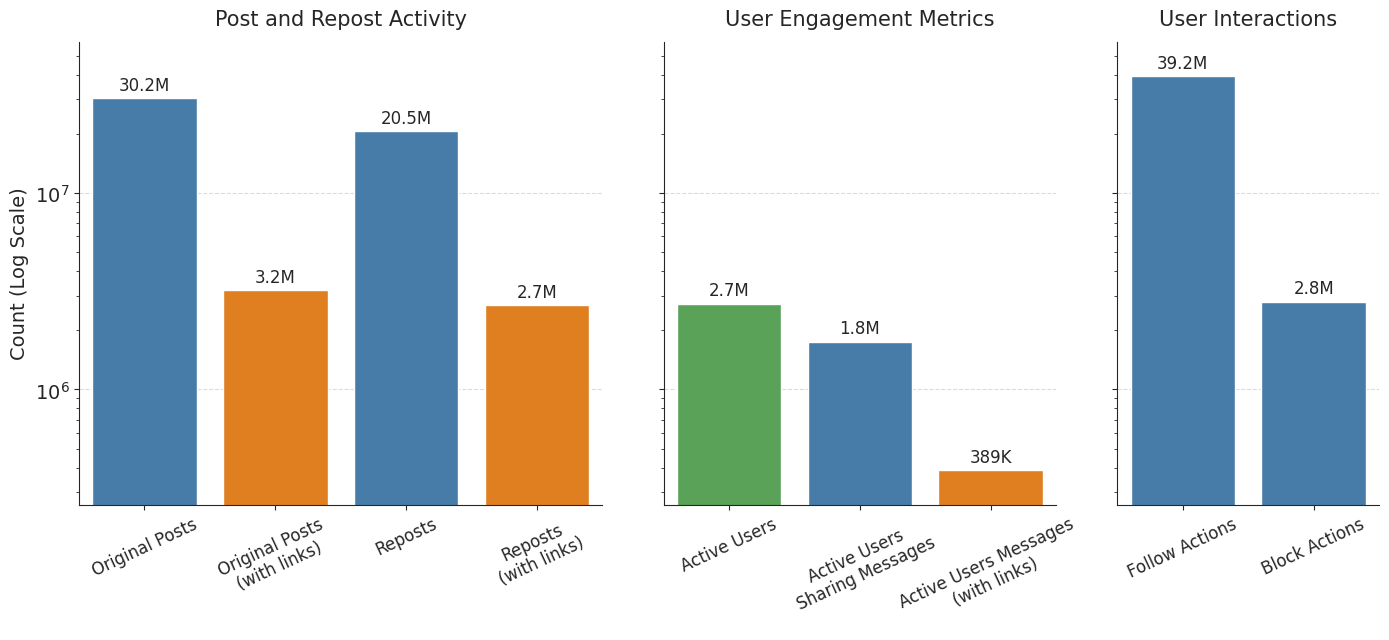}}
\caption{Bar chart summarizing key dataset statistics over the whole observation period: total posts, posts containing links, active users, shared messages (with links), follow actions, and block actions. Values are presented on a logarithmic scale due to different orders of magnitude.}
\label{fig:characteristics}
\end{figure}

Bluesky enables the tracking of various user activities on the platform. Our analysis focuses on key actions: posts, replies, reposts, follows, and blocks, which constitute the primary forms of user interaction and communication. Users follow others to stay updated, create posts to share content, repost content from others, reply to engage in discussions and block users to prevents another users from interacting with them.

While Bluesky's terms of service do not impose privacy restrictions on data collection, we strictly collect only publicly available user information, posts, and associated metadata in accordance with its Privacy Policy.\footnote{\url{https://bsky.social/about/support/privacy-policy}} We do not publicly release the collected data and provide only anonymized information in this paper, except for select prominent accounts discussed in \S\ref{sec:4.8}.

Fig.~\ref{fig:characteristics} presents key statistics of our dataset, collected over a 56-day period from January 9 to March 4, 2024 and encompassing 114 million user activities, including posting activities as well as follow and block actions.
To analyze the language distribution in shared content, we first removed irrelevant content (e.g., URLs, emojis, etc.) and then applied a language classifier using the \texttt{langdetect}\footnote{\url{https://pypi.org/project/langdetect/}} NLP library.

\subsection{News source labeling}
\label{sec:labeling}

To evaluate the reliability of news outlets shared on Bluesky, we label web domains using NewsGuard\footnote{\url{https://www.newsguardtech.com/}} ratings, following a well-established approach in the literature~\cite{cheng2021twittervsfacebook,pierri2023one,Lühring2025BestPractices}. NewsGuard is a reputable, independent organization that employs experts to assess news sources based on criteria such as transparency, accountability, adherence to journalistic standards, and error correction. Its ratings range from 0 (highly unreliable) to 100 (highly reliable), providing a standardized measure of news source credibility. Following prior work~\cite{Nogara2024Misinfo,pierri2020diffusion,pierri2021vaccinitaly}, we classify news outlets with a NewsGuard rating of 30 or lower as \textit{low-credibility}, and those with a rating higher than 60 as \textit{mainstream} news sources.

We also leveraged political bias ratings from Media Bias/Fact Check (MBFC),\footnote{\url{https://mediabiasfactcheck.com/}} an independent organization that evaluates news media sources, to classify the political leaning of news websites shared on Bluesky. Sources are categorized along a seven-point political spectrum: Extreme Left (-3), Left (-2), Left-Center (-1), Least Biased (0), Right-Center (1), Right (2), and Extreme Right (3). We compute the political leaning score of a user by averaging the leaning of the sources shared by that user across their posts, following previous work~\cite{Cinelli2021EchoChamber}. 

Approximately 6.2 million posts (8.7\% of all posts) contained a URL. Among these, we rated about one million posts using NewsGuard scores, representing 16\% of posts sharing URLs, and approximately 850k posts with MBFC political ratings, representing 13.6\% of posts containing URLs.

\subsection{Community detection}
\label{sec:community}

To study user communities and interactions, we build a directed, weighted network based on user reshare activities. Users are represented as nodes and an edge ($i \rightarrow j, w$) represents user $j$ resharing content by user $i$ $w$ times. 
We apply the Louvain community detection algorithm~\cite{Louvain} to the undirected, unweighted version of this graph, obtaining 14,381 different communities. We studied the five largest communities, which represent 87\% of users in our data. 
To gain a deeper understanding of the discussions within each community, we manually examined both users and the content they shared.
For each community, we selected a sample of 30 users that included both randomly chosen members and those with the highest sum of in- and out-degree. 
We analyzed the posts shared by these users and, when possible, inspected their active profiles on the Bluesky web interface to observe their most recent activity and interactions.

\section{Results}

\begin{figure}[!t]
\centerline{\includegraphics[width=\linewidth]{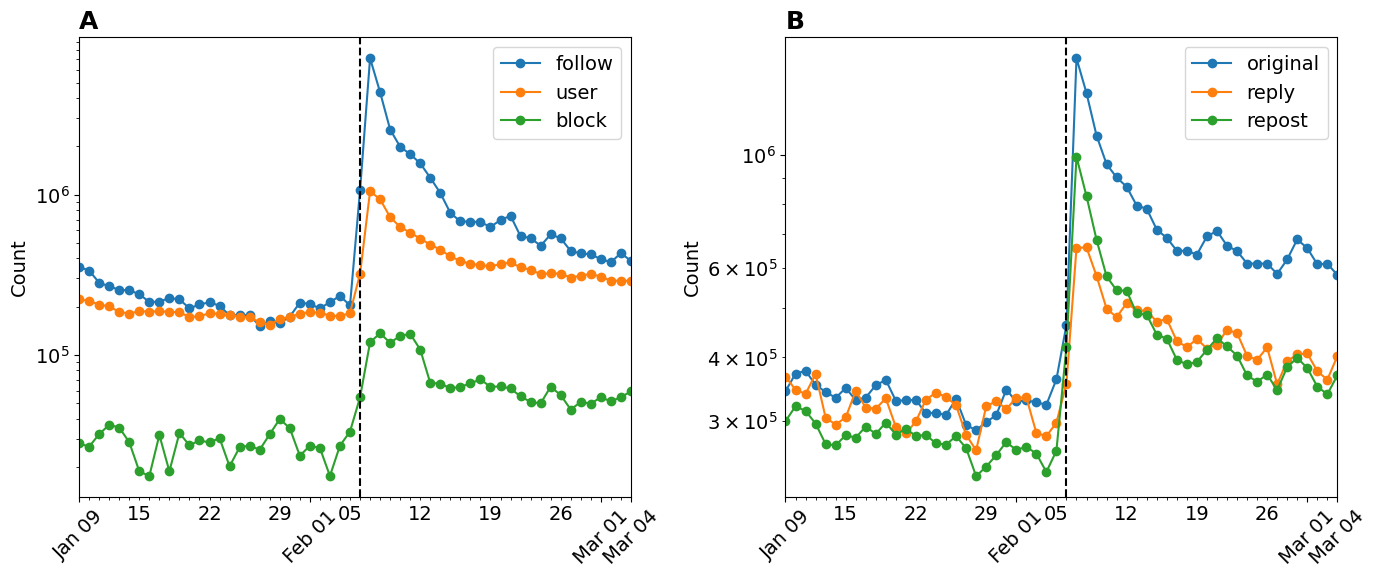}}
\caption{Online activity on Bluesky before and
after the public opening (Feb. 6th), indicated by the dashed line.}
\label{fig:before-after-timeseries}
\end{figure}

\subsection{Temporal patterns of online activity}

As shown in Fig.~\ref{fig:characteristics}, original posts are the most common user activity on the platform, suggesting that users prefer creating new content over resharing or interacting with existing posts. This trend may be driven by the platform's rapid user-base expansion and it contrasts with existing centralized social media platforms, such as Twitter/X, where resharing via retweets is more prevalent~\cite{alshaabi2021growing,luceri2021down}.

Fig.~\ref{fig:before-after-timeseries}A presents the daily number of unique active users along with follow and block actions. 
The platform's public opening on February 6th (dashed line) triggered sharp spikes in activity, peaking at one million active users and over 7 million follow actions the next day. These surges represent a nearly six-fold increase in active users and a 35-fold increase in follow actions compared to the previous day. Both trends declined rapidly in the following days, eventually stabilizing at slightly higher levels than those observed before the opening, likely reflecting the waning initial excitement around the platform.  

The rise in users and follow actions was accompanied by a 3.6-fold surge in blocking activity, increasing from 33,269 to 120,054 instances. Blocking activity stabilized after a few days but remained slightly elevated compared to pre-February 6 levels. During the observation period, 287,539 users blocked a total of 758,681 accounts, accounting for 2.7 block actions received per user on average. 
However, due to the heterogeneous nature of blocking behavior, with some users blocking a disproportionately large number of accounts, the mean number of accounts blocked per user was higher (9.5).

Fig.~\ref{fig:before-after-timeseries}B depicts the volume of shared posts over time, categorizing user activities into original posts, replies, and reposts. As previously noted in Fig.~\ref{fig:characteristics}, original posts are the most prevalent form of shared activity, a trend that becomes even more pronounced following the platform's public launch.  

As expected, the platform's opening on February 6th triggered a sharp increase in shared content across all types of posts, driven by the influx of new users. The volume of original posts surged 4.3-fold, from 362k on February 5th to 1.5M the next day. Reposts and replies also saw substantial growth, rising from 262,201 and 297,717 to 990,835 and 656,063, respectively. Despite these initial spikes, posting activity rapidly declined in the following days.

\subsection{Prevalence of different languages}
\label{language_analysis}

\begin{figure}[!t]
\centerline{\includegraphics[width=\linewidth]{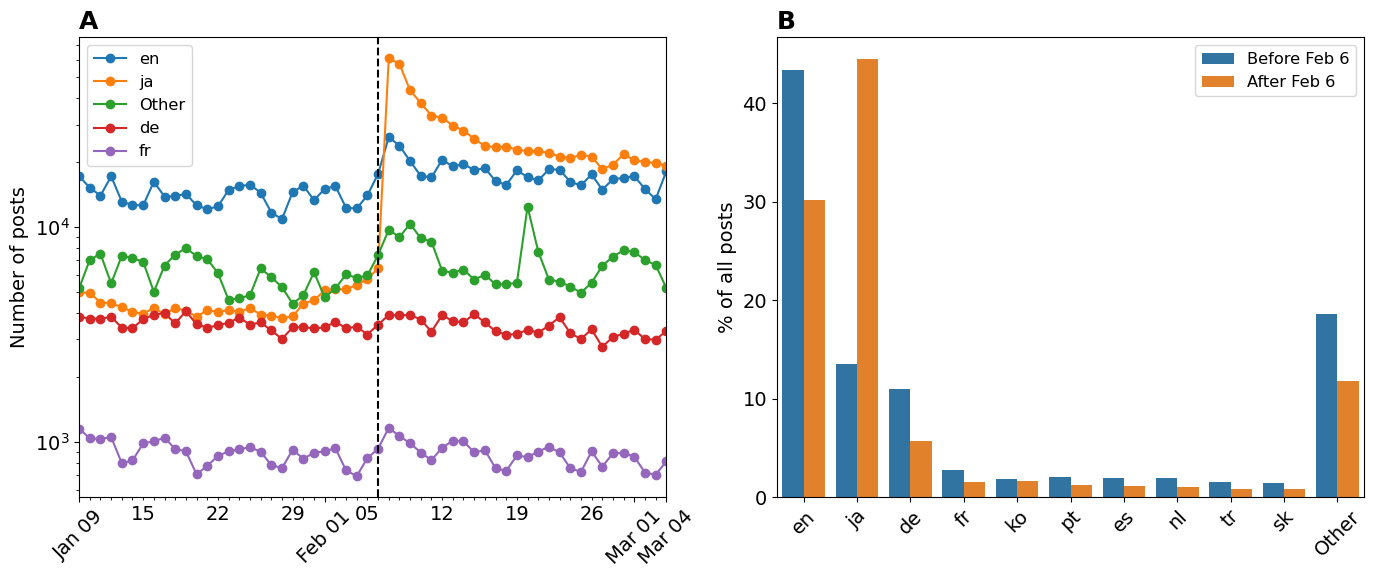}}
\caption{\textbf{(A)} Trend of 5 top languages on Bluesky during the observation period. \textbf{(B)}~Top 10 languages in Bluesky. Bars of the same color sum to 100\%.}
\label{fig:language}
\end{figure}

Fig.~\ref{fig:language}A shows the trends of the top five languages used throughout the observation period. Japanese posts surged immediately after the platform's opening, becoming the most used language. In contrast, content in other languages, including English, remained relatively stable, exhibiting only minor fluctuations following the opening.

Fig.~\ref{fig:language}B shows the prevalence of the top 10 languages in user posts, highlighting the dominance of English and Japanese, which together account for more than two-thirds of all content. The share of English content declined from 43\% to 30\% following the platform's opening, while Japanese content increased from 14\% to 44\% over the same period.

\subsection{Changes in the follower network}
\label{sec:4.9}

Similarly to what we did for the reshare network, we build another network using the follow actions in the dataset. In this network, a directed edge ($i \rightarrow j$) represents user $i$ following user $j$. 
The sharp increase in \textit{follow} activities after the public launch, shown in Fig.~\ref{fig:before-after-timeseries}, is reflected in various follower network statistics measured before and after the platform's opening, as detailed in Table~\ref{table:network-statistics}.  

\begin{figure}[!t]
\centering
\includegraphics[width=1\linewidth]{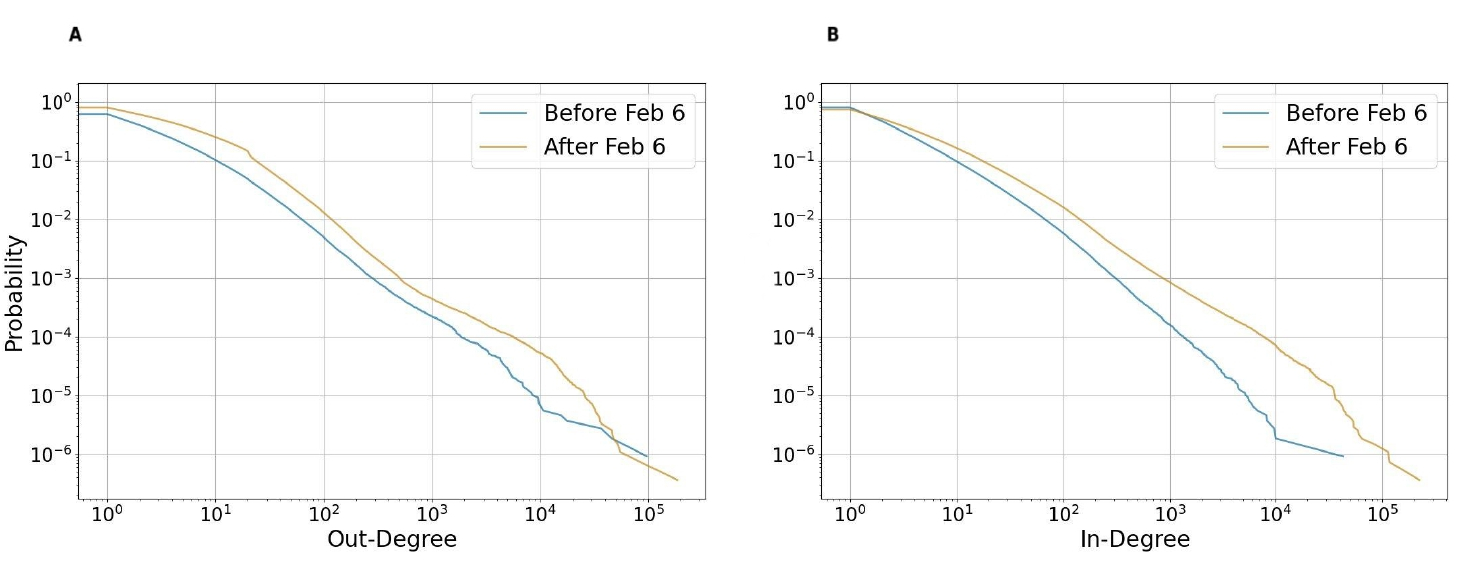}
\caption{Complementary cumulative distributions of node \textbf{(A)} out-degree and \textbf{(B)}~in-degree in the follower network.}
\label{fig:in-out-degree}
\end{figure}

While the density of the follower network slightly decreased post-opening, the size of the strongly connected component more than tripled, and the average degree more than doubled. 
This suggests that Bluesky users tend to follow more accounts after the opening.
The out-degree distributions in Fig.~\ref{fig:in-out-degree}A further confirm this trend.

\setlength{\tabcolsep}{12pt}
\begin{table}[t]
\caption{Follower network statistics. LSCC stands for the largest strongly connected component.}
\centering
\label{table:followers}
\begin{tabular}{lccc}
\cline{2-4}
                 & Before Feb. 6 & After Feb. 6  & Difference      \\ \hline
Number 			of nodes & 1,088,539 & 2,751,272  &   1,662,733       \\
Number 			of edges & 5,230,054 & 28,838,739  & 23,608,685         \\  
Density & $4.4 \times 10^{-6}$    &  $3.8 \times 10^{-6}$ & $-0.6 \times 10^{-6}$\\
Avg. in-degree       & 9.6     & 20.3 &   10.7           \\
Avg. out-degree       & 7.4     & 14.9 &   7.5           \\
LSCC size      & $\sim$200k  & $\sim$650k &     $\sim$450k       \\
\hline
\end{tabular}
\label{table:network-statistics}
\end{table}
\setlength{\tabcolsep}{6pt}

The average number of followers per user increased from 9.6 before February 6th to 20.3 after, while the median rose from 3.0 to 4.0. Similarly, the average number of accounts followed increased from 7.4 to 14.9, with the median rising from 2.0 to 4.0.  
Fig.~\ref{fig:in-out-degree} plots the distributions of out-degree (number of follow actions) and in-degree (number of followers). 
The ten accounts that gained the most followers and the ten accounts that followed the most users before and after the opening of the platform are presented in Table~\ref{table:follower-followee}.

\begin{table}[t]
\caption{Annotated list of the ten users with the most new followers and the ten users who performed most follow activities before and after the platform's opening. Only the real names of prominent accounts are included for privacy reasons. `JP' indicates Japanese-speaking users.}
\centerline{\footnotesize
\begin{tabular}{lr|lr||lr|lr}
\hline
\multicolumn{4}{c||}{Accounts with most new followers} &\multicolumn{4}{c}{Accounts who followed most accounts}     \\ \hline 
\multicolumn{2}{c|}{Before Feb. 6} & \multicolumn{2}{c||}{After Feb. 6} &  \multicolumn{2}{c|}{Before Feb. 6} & \multicolumn{2}{c}{After Feb. 6} \\ \hline 
Account    & Num.    & Account    & Num.   &   Account    & Num.    & Account    & Num.        \\ \hline
Bluesky & 43,801& Bluesky  & 120,709  & user & 50,570 & user  & 167,220   \\   
user  & 10,256 & user JP &88,438&user&24,588& user JP&85,075\\
user &    9,886      & user &66,535&user &23,482&user JP &55,073\\
Wash. Post &8,508&NY Times &62,087&user &21,907& user JP &45,848\\
NY Times &8,393&Wash. Post &60,904&user &13,414&user  &44,588\\
user&6,809&user&55,081&user&12,310&user &44,460\\
Bluesky CEO&6,314&user JP&54,944&user&10,689&user  &43,786\\
user &5,744&Bloomberg &54,800&user&9,950&user &40,926\\
user&5,621&user&52,455&user&8,211& user JP&37,209\\
Bloomberg &5,332&user JP&49,697& user&7,588& user JP&36,189\\ \hline
\end{tabular}
}
\label{table:follower-followee}
\end{table}

This analysis suggests potential spamming or automated behavior, as some accounts followed an unusually large number of users immediately after joining Bluesky (Fig.~\ref{fig:in-out-degree}A).
The crossover in the tails of the out-degree CCDF curves (just before out-degree $10^{5}$) results from the uneven distribution of follow activities and the disparity in the number of users before and after the platform's opening.
Notably, in each period, a single outlier user performed an exceptionally high number of follow actions --- 50,570 before and 167,220 after, approximately 2--3 times the number of actions of other users listed in Table~\ref{table:follower-followee}.  
Since the number of users performing follow actions before the opening is about one-third of those after (see Table~\ref{table:network-statistics}), the probability in the CCDF tail is higher for the earlier group. When these two outlier users are removed, the crossover between the CCDF curves disappears.

Among the users who gained the most followers (Table~\ref{table:follower-followee}), several remained consistent across both periods, with the majority being news outlets such as \textit{The Washington Post}, \textit{The New York Times}, and \textit{Bloomberg}. This prevalence of news organizations among the most-followed accounts suggests that Bluesky may be evolving into a platform for news dissemination, potentially positioning itself as a replacement for Twitter.  
Despite this trend, three new Japanese accounts entered the top ten following the platform's public launch, with one rising to second place---surpassed only by the official Bluesky account.

Before the platform's opening, the users who followed the most new accounts were primarily English-language accounts and appeared to engage in normal activity, with the exception of one user who was suspended by Bluesky.  
After the opening, however, the composition of these users changed significantly: half were Japanese, two accounts were deleted, one was suspended, and two were flagged as spammers by Bluesky. Unlike the overall network behavior, these users were more engaged in reposting activities than in creating original posts.

\subsection{Political leaning of Bluesky users}

\begin{figure}[!t]
\centerline{\includegraphics[width=\linewidth]{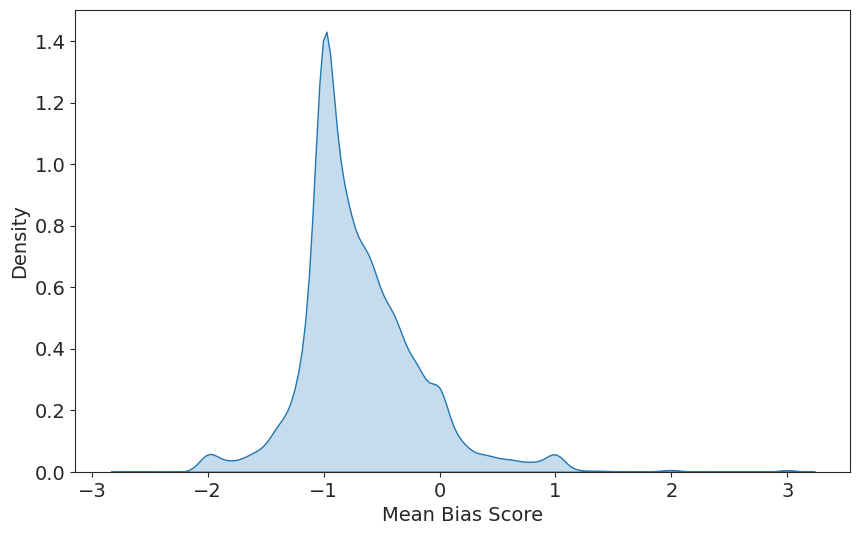}}
\caption{Distribution of the average political leaning score of users that shared at least 5 links to rated domains.
}
\label{fig:political}
\end{figure}

Fig.~\ref{fig:political} presents the distribution of estimated political alignment among active Bluesky users, defined as those who shared at least five posts linking to rated websites. Each user's political alignment is determined by averaging the political bias scores of the websites they shared during the observation period (see \S\ref{sec:labeling}). We can observe that the distribution is skewed towards liberal leaning, consistent with previous findings reported for Twitter before Musk's acquisition.\footnote{\url{https://www.pewresearch.org/internet/2019/04/24/sizing-up-twitter-users/}} We present the distribution of political leanings computed over the entire observation period, as no significant differences were detected before and after the platform's opening (Mann-Whitney U test, $p = 0.05$).

We determine each user's political leaning based on the average bias score of their posts. In this analysis, we make the simplifying assumption that a negative bias score indicates a left-leaning, a score of zero indicates a centrist, and a positive score indicates a right-leaning, acknowledging that this thresholding is inherently arbitrary. The analysis revealed that the majority of active users on the platform are classified as left-leaning (bias score $< 0$), accounting for $74.61\%$ of the sample. In contrast, $18.17\%$ of users are categorized as centrist (bias score $= 0$), and only $7.21\%$ are identified as right-leaning (bias score $> 0$). Accordingly, we report the distribution over the entire observation period, as the comparison between the pre- and post-opening phases of the platform did not yield a significant difference.

\subsection{Credibility of information shared on Bluesky}

Overall, the prevalence of low-credibility content shared on Bluesky during the period of analysis is negligible, comprising only 0.08\% of all posts. Similarly, the proportion of users who shared at least one link to low-credibility sources is very small, accounting for just 0.13\% of all users.

\begin{figure}[!t]
\centerline{\includegraphics[width=1\linewidth]{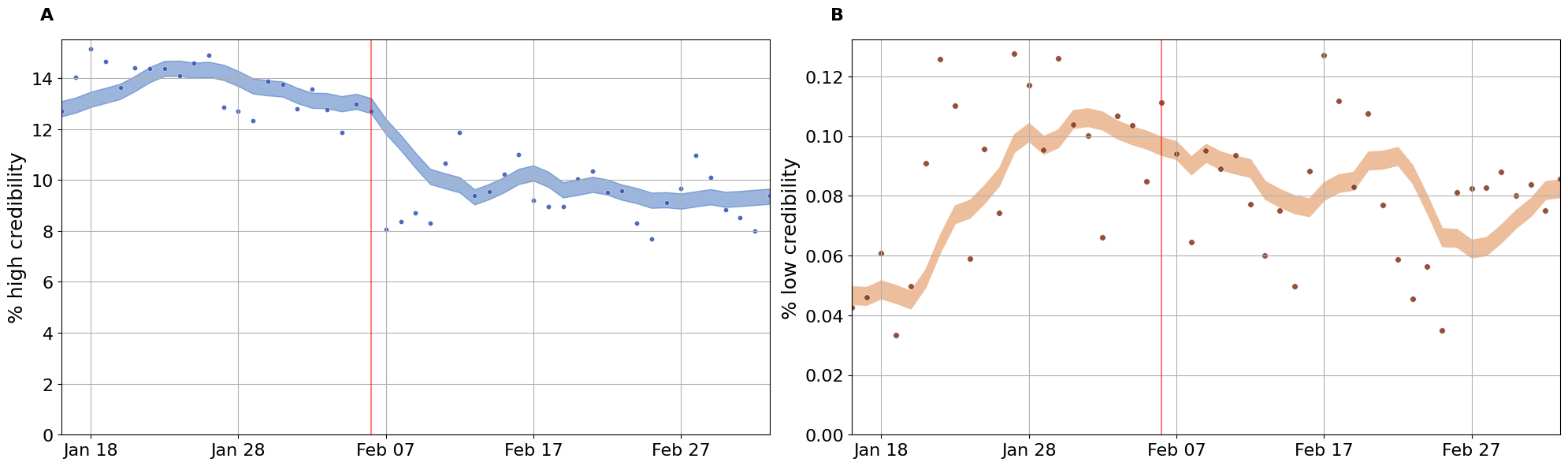}}
\caption{Moving average (7-day) of the percentage of high credibility and low-credibility domains shared by users. \textbf{(A)} Percentage of posts sharing links to high-credibility domains, i.e., NewsGuard ratings \(\geq 60\). \textbf{(B)} Percentage of posts sharing links to low-credibility domains, i.e., NewsGuard ratings \(\leq 30\)}
\label{fig:credibility-domains}
\end{figure}

Fig.~\ref{fig:credibility-domains} presents the daily proportion of posts sharing links to high- and low-credibility websites. We observe a noticeable decline in the share of high-credibility domains after the platform's public opening, whereas the proportion of low-credibility domains remains relatively stable throughout the observation period. 
Despite these trends, high-credibility domains are shared 125 times more frequently than low-credibility ones on an average day.
The median daily share rate is approximately 10\% for high-credibility domains, compared to just 0.08\% for low-credibility domains.

\begin{figure}[!t]
\centerline{\includegraphics[width=1\linewidth]{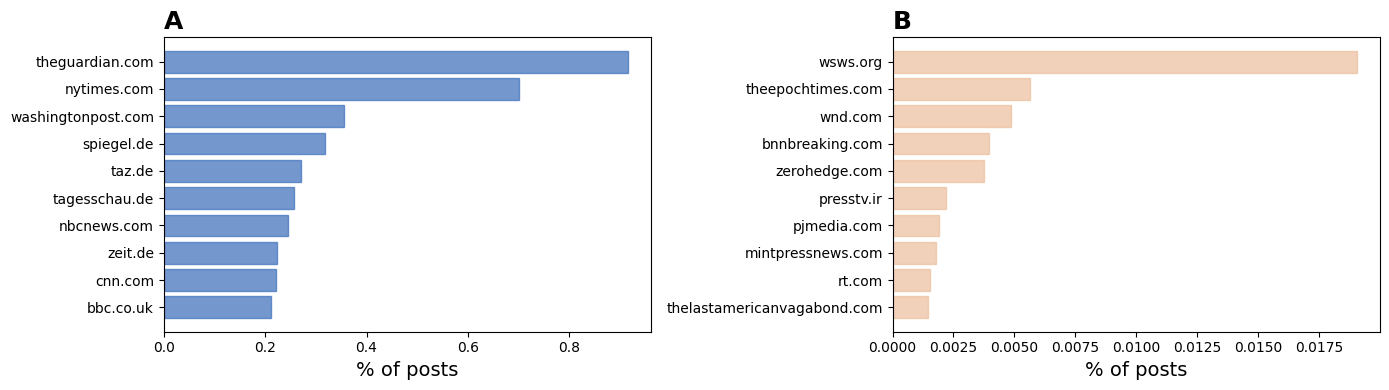}}
\caption{Most shared \textbf{(A)} high-credibility and \textbf{(B)} low-credibility websites during the period of analysis.}
\label{fig:credibility-websites}
\end{figure}

Figure~\ref{fig:credibility-websites} presents the most frequently shared low- and high-credibility websites throughout the entire observation period. The composition of low-credibility domains remained largely consistent before and after Bluesky’s public opening, with one notable exception: \url{wsws.org}, a far-left socialist site known for publishing disputed claims, which saw its share decrease by half compared to pre-opening levels. However, this decline is not due to a reduction in the actual sharing of this domain but rather to the overall increase in the sharing of all domains. As expected, many of these low-credibility sources are news agencies that have been previously accused of spreading disinformation~\cite{Hellman2024}. Regarding high-credibility domains, a significant observation is the presence of German news agencies---e.g., \url{spiegel.de} and \url{taz.de}---alongside English-language sources, such as \url{nytimes.com}, and \url{theguardian.com}.
This suggests that German users may be more active on Bluesky compared to other social networks, potentially reflecting regional differences in platform adoption.

Consistent with findings reported for other social media platforms~\cite{DeVerna2022FIB,Nogara2022TheDD,pierri2021vaccinitaly,Grinberg2019Jan}, we observe evidence of so-called ''superspreaders,'' i.e., a small subset of users responsible for the majority of unreliable content shared on Bluesky. Specifically, we find that ten accounts (1.8\% of all users who shared low-credibility content) were responsible for disseminating 62\% of links to low-credibility sources. 

\subsection{Toxicity of Bluesky conversations}

We analyzed the toxicity of online conversations on Bluesky using Detoxify~\cite{Detoxify}, a model used to detect toxic comments, focusing on the languages supported by the model (English, Italian, French, Russian, Portuguese, Spanish, and Turkish). We present results for the entire observation period, as no significant temporal variations were detected in toxicity after the public opening.

\begin{figure}[!t]
\centerline{\includegraphics[width=1\linewidth]{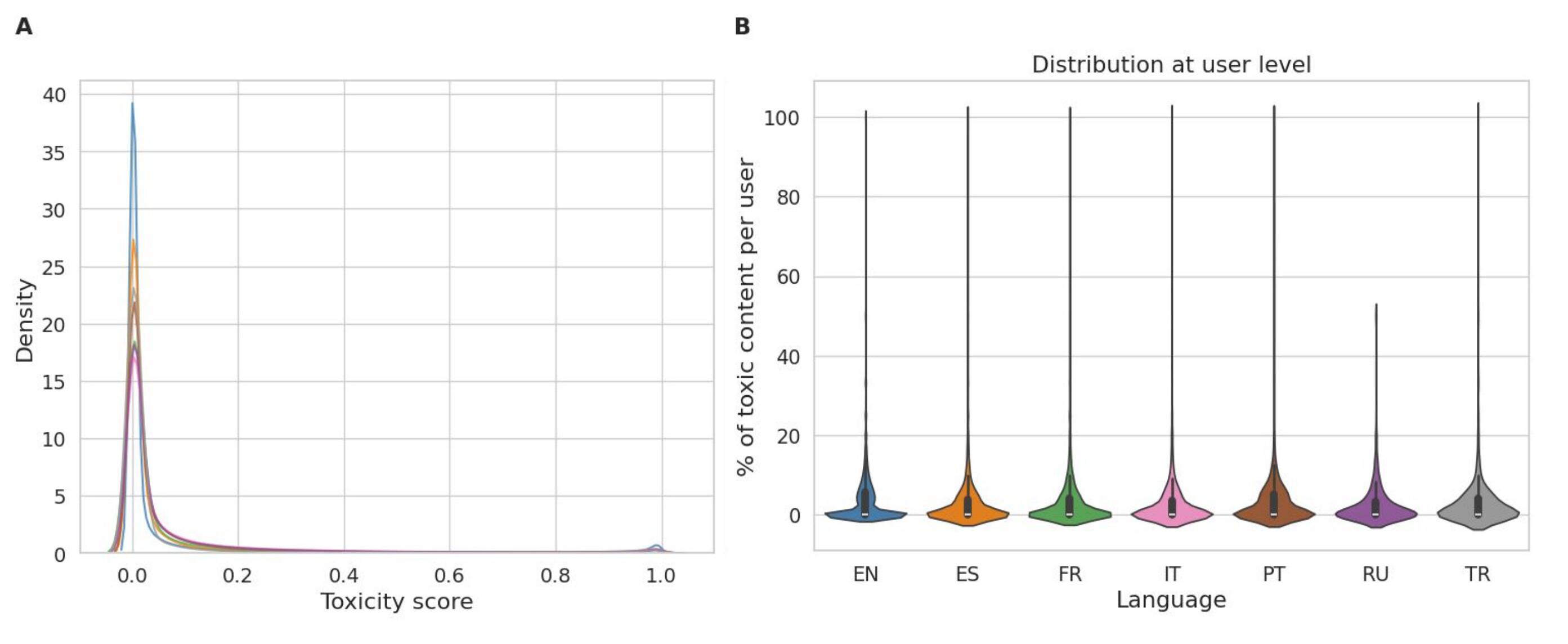}}
\caption{
Toxicity of Bluesky posts. \textbf{(A)} Density distribution of toxicity for the different languages. The color assignment is indicated in the companion figure. 
\textbf{(B)} Distributions of the percentages of posts containing toxic content per user, by language.
}
\label{fig:toxicity}
\end{figure}

While the the distributions of toxicity scores are different across different languages ($p<0.05$ according to a Kruskal-Wallis test), Fig.~\ref{fig:toxicity}A shows that only a small fraction of posts exhibit concerning levels of toxicity in all cases.
The same pattern holds for user toxicity, computed by averaging the toxicity scores of all posts each user shared, as shown in Fig.~\ref{fig:toxicity}B. In this analysis we did not use a threshold, but results are similar when we eliminate users with only one post (less than 7\%). 

After a Bonferroni correction, we collected the pairs of languages with significant differences in toxicity per user ($p<0.05$): English (EN) differs from Spanish (ES), Russian (RU), Italian (IT), and French (FR); and Portuguese (PT) differs from Spanish (ES), Russian (RU), Italian (IT), and French (FR).

\subsection{Interplay between misinformation, political leaning, and toxicity} 
\label{sec:4.8}

We further examined the interplay between misinformation, political leaning, and toxicity for Bluesky users over the entire period of analysis.
To explore whether these factors varied with users' activity levels---for example, whether more active users were more likely to share toxic or misleading content---we divided users into categories based on how frequently they posted.
Users who posted between one and nine times were considered \textit{low activity}, those with 10 to 99 posts were classified as \textit{high activity}, and users with 100 or more posts were classified into a \textit{hyper activity} group.

\begin{figure}[!t]
{\centerline{\includegraphics[width=1\linewidth]{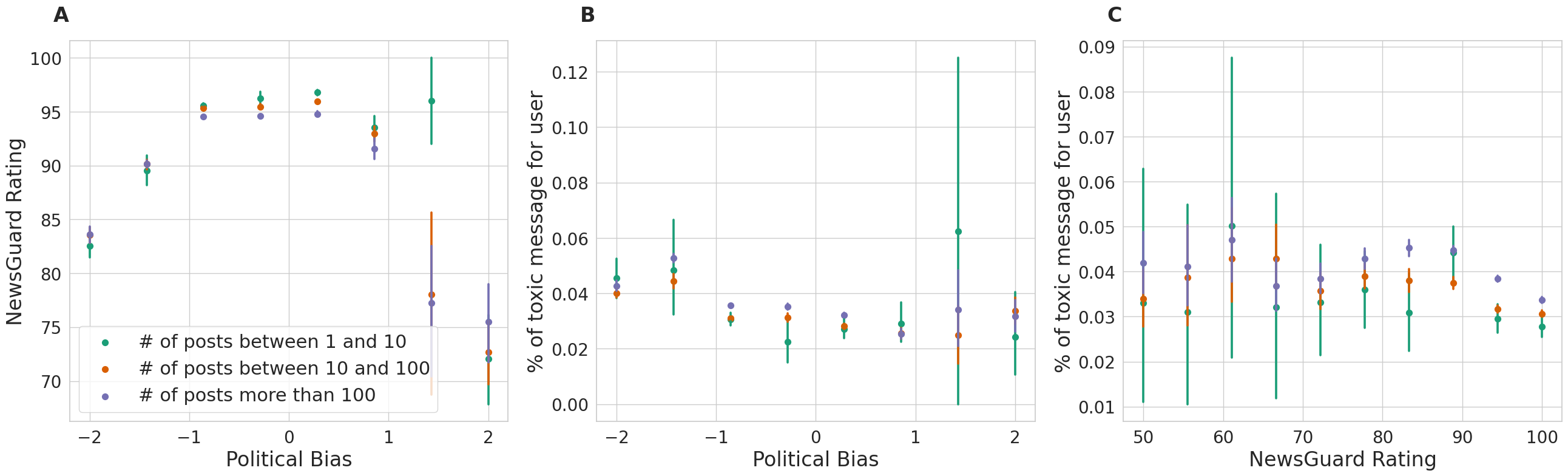}}
}
\caption{Interplay between metrics by user activity levels. The error bars represent the standard deviation. \textbf{(A)} Relationship between political bias and reliability, \textbf{(B)} between political bias and toxicity, and \textbf{(C)} between reliability and toxicity. The NewsGuard rating in the panel (C) is depicted only from 50 to 100 (even if NewsGuard rating goes from 0 to 100) as no ratings below 53 were present.}
\label{fig:interplay-newsguard_bias}
\end{figure}

 
As shown in Fig.~\ref{fig:interplay-newsguard_bias}A, we observe that extreme users, on the left and the right side of the political spectrum, are associated with a lower quality of shared information (computed as the average rating of all the links they shared) compared to other users. We do not find significant differences between the three classes of users (low, high, and hyper active users), indicating that these patterns hold regardless the number of user posts.

Fig.~\ref{fig:interplay-newsguard_bias}B illustrates the correlation between toxic behavior and political leaning. We observe that left-leaning users exhibit higher toxicity levels than center or right-leaning users, which may be attributed to their greater presence on the platform. This behavior is particularly pronounced among more active users.

Finally, as shown in Fig.~\ref{fig:interplay-newsguard_bias}C, we do not observe significant patterns in the relationship between toxic behavior and misinformation sharing on the platform. 

\subsection{Analysis of communities}
\label{sec:4.10}


We now explore the five largest reposting communities identified using Louvain's algorithm (see \S \ref{sec:community} for details).
These communities are numbered by size, with community 1 being the largest and community 5 the smallest.
They are relatively homogeneous, organized primarily by language groups and type of content shared. 
Table~\ref{table:communities} shows the percentages of users in each community.

\begin{table}[t]
\caption{Annotated list of the five largest user-level communities with the respective percentages of users in the dataset, the dominant language of the community, and a brief description of the content shared by their users.}
\centering
\begin{tabular}{c|c|c|l}
\hline
Community & \% of users & Language & Description             \\ \hline
1                & 35.5        & Ja       & explicit art cartoons \\
2                & 19.7        & En       & democrats   \\
3                & 13.7        & En       & art cartoons            \\
4                & 10.4        & En       & explicit art cartoons \\
5                & 7.5         & De       & democrats \\
\hline
\end{tabular}
\label{table:communities}
\end{table}

Through a manual review of hashtags, linked domains, and images (see \S \ref{sec:community} for details), we found that much of the content circulating within these communities is related to artistic and cartoon themes.
A strong Japanese-language presence is evident in the dataset, particularly within community 1, which is characterized by apolitical, artistic, and explicit content. 
Similar themes appear in communities 3 and 4, which also share English-language artistic material, though community 4 is more explicit in tone and subject matter.
In contrast, communities 2 and 5 are politically focused, with English and German as their dominant languages respectively. Reviewing shared domains within these communities reveals that both exhibit a left-leaning political orientation.



\begin{figure}
\centerline{\includegraphics[width=1\linewidth]{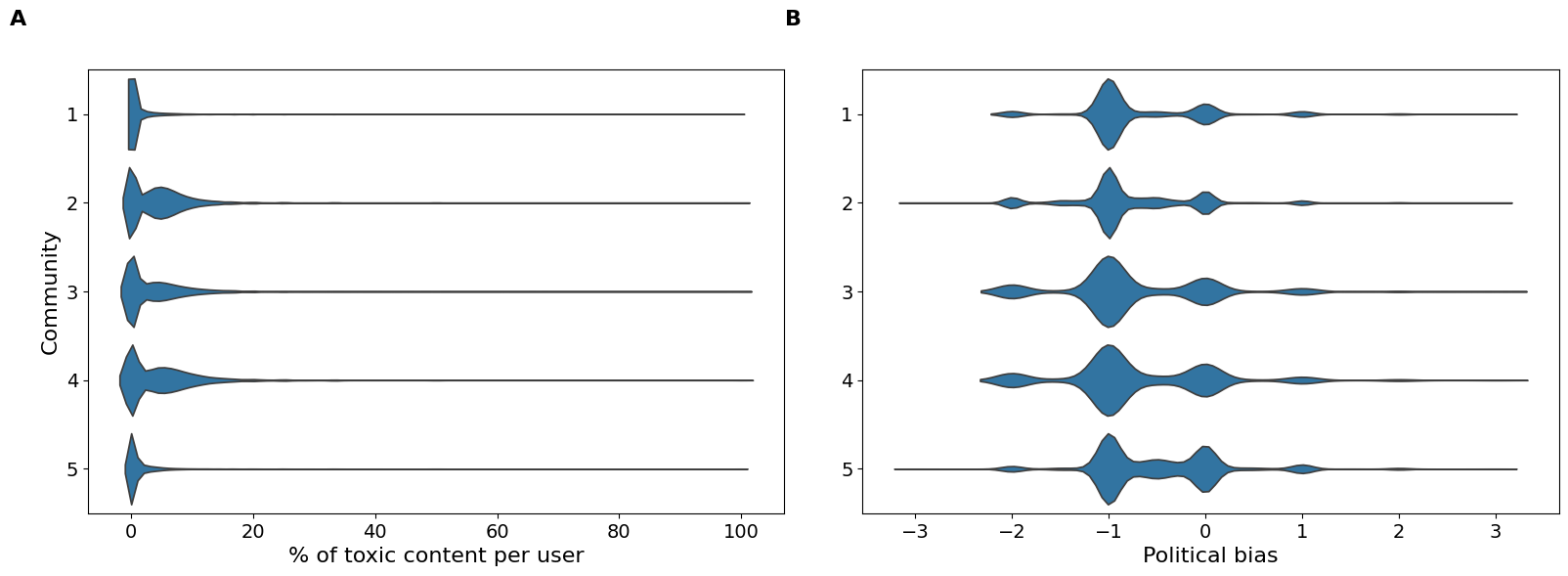}}
\caption{User-level community analysis. \textbf{(A)} Distributions of percentages of toxic content. \textbf{(B)} Distributions of political bias, where -3 means extreme left, 0 center, and 3 extreme right.}
\label{fig:communities-analysis}
\end{figure}

As shown in Fig.~\ref{fig:communities-analysis}A, we computed the distributions of toxic messages shared on average by users across different communities. We used a toxicity score threshold of 0.5 to define a toxic text~\cite{hua_2020_towards, saveski_2021_the}. The analysis of the distributions indicates that communities 2 and 4 exhibit a higher proportion of users engaging in the dissemination of toxic messages, with median values of 3.7\% and 3.5\% toxic content per user, respectively. In contrast, community 3 demonstrates lower levels of toxicity, characterized by a median value of 1.0\%. Users within communities 1 and 5 display minimal to negligible toxic behavior relative to other communities, both presenting a median toxicity level of 0.0\%. While the toxicity does not vary by community topic, it does vary by language, with English having a higher level of toxicity than German and Japanese. 
Kruskal-Wallis tests with Bonferroni correction reveal statistically significant differences ($p<0.001$) among all community pairs except for between communities 2 and 4.

We also analyzed the distribution of political leanings among users in different communities. As shown in Fig.~\ref{fig:communities-analysis}B, the overall political orientation skews left, with community medians clustering between -0.67 and -1, in line with the general Bluesky population.  


\subsection{Content moderation}
\label{sec:4.11}

We further examined moderation on the platform by analyzing the status of user accounts.
To do this, we used the \texttt{getProfiles} endpoint,\footnote{\url{https://docs.bsky.app/docs/api/app-bsky-actor-get-profiles}} which provides information about user profiles, including their status. 

A user's account status varies depending on their standing on the platform.
Active accounts return a status of \emph{Online}, while deleted accounts return \emph{InvalidRequest}, indicating that the profile was likely canceled by the user. 
Moderated accounts produce one of two responses---\emph{AccountDeactivated} or \emph{AccountTakedown}---both signaling a policy violation. The former typically indicates a temporary enforcement action, while the latter reflects a more permanent removal.

As of November 2024, most users in our dataset (95.9\%) remained active.
In contrast, 3.6\% of accounts had been deleted, 0.4\% were temporarily deactivated, and 0.1\% had been permanently removed---indicating that 0.5\% of users had been subject to platform moderation.

\begin{figure}
 \centering\includegraphics[width=1\columnwidth]{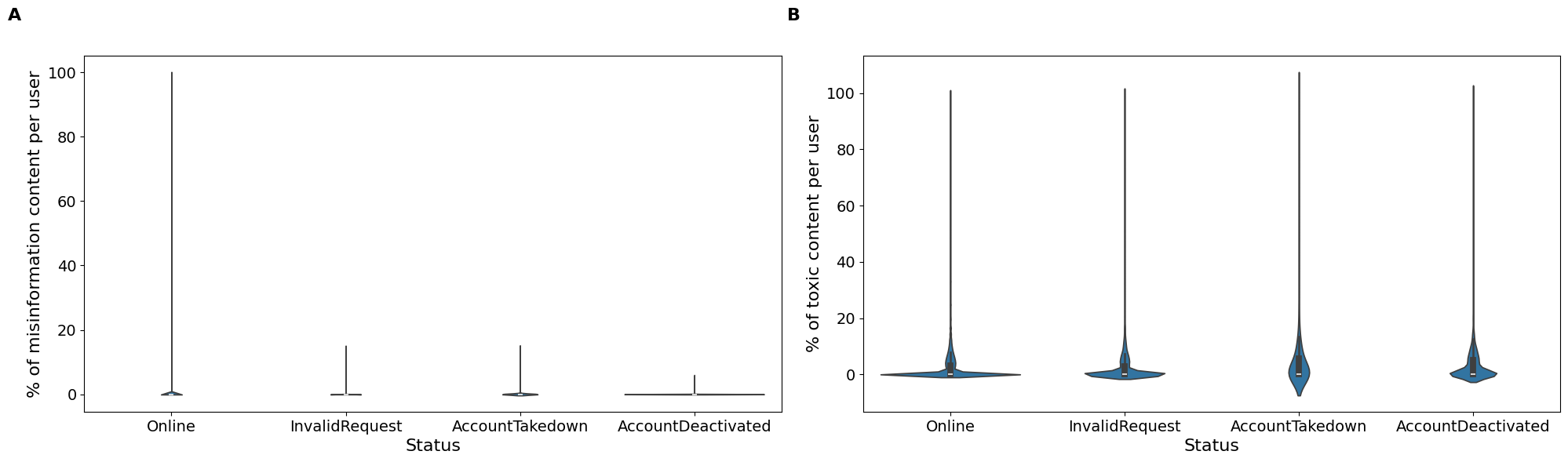}
\caption{Analysis of user status categories. \textbf{(A)} Percentage of misinformation shared. \textbf{(B)} Percentage of toxic content shared.}
\label{fig:moderation-metrics}
\end{figure}

To understand the reasons for user moderation, we analyzed the extent to which users in each of these groups shared toxic and low-credibility content. 
Fig.~\ref{fig:moderation-metrics}A suggests that moderation was done against users who violated the terms of service, as users who shared links from low-credibility sources remain. 
Similarly, in the case of toxicity, the moderated categories do not exhibit different behavior from other users, as shown in Fig.~\ref{fig:moderation-metrics}B, likely because toxicity is not officially mentioned in the community guidelines~\cite{community}. 
Kruskal-Wallis tests with Bonferroni correction were performed on the category of account according to toxicity and misinformation per user.
The tests reveal statistically significant differences ($p<0.001$) between all categories except the pairs \emph{AccountDeactivated}-emph{AccountTakedown} in both cases and \emph{AccountDeactivated}-emph{InvalidRequest} in the case of misinformation.

\section{Discussion}

In this study, we provided the first large-scale analysis of how Bluesky’s transition from an invite-only platform to a publicly accessible one affected user activity and network structure.
Our findings reveal that while Bluesky exhibits a broad distribution of activity similar to more established social media platforms, it stands out due to a higher volume of original content compared to reshared posts and low levels of toxicity. After the public opening, Bluesky experienced a surge in new activity. The influx of Japanese users is an intriguing trend that warrants further investigation, as it may indicate that Bluesky's conversational dynamics are particularly appealing to this demographic.

We uncovered evidence of suspicious user behavior, including accounts that followed large numbers of users and shared content from low-credibility news outlets. A small subset of users was responsible for the majority of unreliable content shared, a pattern previously observed on other platforms like Twitter/X. Following the platform’s public opening, some of the most aggressive account followers were flagged as spam or even deleted or suspended, suggesting attempts to misuse the platform. The fact that some of these suspicious actors were swiftly banned or labeled as spam indicates that content moderation mechanisms are actively functioning on Bluesky. Our analysis of account statuses further reveals that only 0.5\% of users have been moderated, while 3.6\% are no longer on the platform, leaving 95.9\% of users active~\cite{broniatowski_2025_explaining}.

We examined the reshare network, identifying that the five largest communities accounted for 87\% of users. Our findings suggest that English-speaking communities exhibit higher levels of toxicity compared to other language-based communities, though overall, toxicity levels remained relatively stable across different languages. We also confirmed that Bluesky users lean predominantly left-wing politically, as indicated by the classification of shared domains and the political orientation of the largest communities.

Our findings have several important implications for understanding decentralized social media dynamics and content moderation. The relatively low levels of toxicity on Bluesky, combined with its high proportion of original content, suggest that decentralized platforms may foster a different type of user engagement compared to centralized counterparts. However, the emergence of suspicious activity, including users sharing low-credibility information and engaging in mass-following behaviors, indicates that Bluesky is not immune to manipulation attempts. The platform’s ongoing moderation efforts, which have already flagged or removed some misbehaving accounts, highlight both the challenges and potential effectiveness of content governance in decentralized environments.

The left-leaning political orientation of Bluesky's user base raises questions about ideological fragmentation and the extent to which decentralized platforms may cater to specific political communities. The significant influx of Japanese users also suggests that Bluesky may be attracting geographically and linguistically distinct user groups, which could influence the platform’s future development and community structure.

This study has several limitations. First, our analysis covers a relatively short period (56 days), during which the platform underwent a significant transformation as it opened to the public. As a result, our findings capture early trends that may continue to evolve over time. Second, approximately 8\% of reposts could not be traced back to their original posts due to the constraints of our data collection period. Third, the credibility and political bias scores we used were available only for a subset of websites, limiting the scope of our misinformation analysis. Finally, our toxicity analysis was restricted to languages supported by Detoxify---namely English, Spanish, French, Italian, Portuguese, Russian, and Turkish---leaving out other languages that may have distinct toxicity patterns.

Future research should extend the analysis over a longer period to assess the long-term evolution of Bluesky's user behavior, content sharing patterns, and moderation effectiveness. A deeper investigation into the rise of Japanese users on the platform could provide insights into regional adoption patterns of decentralized social media. In addition, the slopes of the two degree distributions should be compared so as to check for any decreases in slope, thus indicating the emergence of larger hubs. Further studies should also explore content moderation approaches in decentralized networks, comparing their effectiveness against those of centralized platforms. Additionally, expanding toxicity and misinformation analysis to a broader range of languages could offer a more comprehensive understanding of harmful content dynamics in multilingual decentralized spaces. Finally, the network analysis could be extended to study block actions among users. 

\section*{Acknowledgments}

This work was partially supported by the Swiss National Science Foundation (grant number CRSII5\_209250) and the Italian Ministry of Education (PRIN PNRR grant CODE prot. P2022AKRZ9 and PRIN grant DEMON \linebreak prot. 2022BAXSPY).

\bibliographystyle{elsarticle-num} 
\bibliography{main}

\end{document}